# Designing a High Efficiency Pulse Width Modulation Step-Down DC/DC Converter for Mobile Phone Applications

**Benlafkih Abdessamad[1], Krit Salah-ddine[2] and Chafik Elidrissi Mohamed[3]**

[1] Laboratory of physic and environment Department of Physique
Faculty of Sciences University Ibn Tofail BP 133
Kénitra 14000, Morocco

[2] Department of informatics Polydisciplinary Faculty of Ouarzazate
University Ibn Zohr, BP/638
Ouarzazate, Morocco

[3] Laboratory of physic and environment Department of Physique
Faculty of Sciences University Ibn Tofail BP 133
Kénitra 14000, Morocco

**Abstract**

This paper presents the design and analysis of a high efficiency, PWM (Pulse-Width-Modulation) Buck converters for mobile phone applications. The steady-state and average-value models for the proposed converter are developed and simulated. A practical design approach which aims at systematizing the procedure for the selection of the control parameters is introduced. The switching losses are reduced by using soft switching, additionally, a simple analog and digital form of the controller for practical realization is provided. It is found that this controller adopts a structure similar to the conventional PWM voltage mode controller. The proposed circuit uses a current-mode control and a voltage-to-pulse converter for the PWM. The circuit, fabricated using a 0.18-μm CMOS technology, reaches a peak load regulation of 20 mV/V and line regulation of 0.5 mV/V at Current load equal 300 mA. The used 10μH inductance and 22μF capacitor and requires clock and Vref/Vramp input of 1,23V.

***Keywords:*** — *High Efficiency, PWM, Buck converters, Soft Switching, CMOS technology.*

## 1. Introduction

Several new techniques for high frequency DC-DC conversion have been proposed to reduce component stresses and switching losses while achieving high power density and improved performance. The key to reducing power consumption while maintaining computational throughput and quality of service is to use such systems at the lowest possible supply voltage. The terminal voltage of the battery used in portable applications (e.g., NiMH, NiCd, and Li-ion) varies considerably depending on the state of their charging condition. For example, a single NiMH battery cell is fully charged to 1.8 V but it drops to 0.9 V before fully discharged [1], [14]. DC–DC converter convert DC voltage signal from high level to low level signal or it can be vice versa depending on the type of converter used in system. Buck converter is one of the most important components of circuit it converts voltage signal from high DC signal to low voltage. In buck converter, a high speed switching devices are placed and the better efficiency of power conversion with the steady state can be achieved. In this paper work performance of buck converter is analyzed. Fig.1.1 shows the basic topology of the converter DC-DC Buck and Fig.1.2 shows a simplified schematic of the buck power stage with a drive circuit block included [2]. and more recently in [8]. With the advent of recent applications with more stringent specifications in terms of regulator settling time, various approximations to time-optimal control have recently been proposed, e.g., in [9], [11], [19], [20] where time-optimal control is described in terms of boundary control and revisited for the buck converter. Reference [12] reexamines the switching-surface time-optimal control and derives analytical equations for the optimal case, similarly to [5]. Recent works in [13], [15], [22], [23] provide a comprehensive account of geometric control principles, including limits of time-optimal control for switching converters. In the field of general control theory, the fundamentals of time-optimal control, which are directly related to the use of Pontryagin's principle, have been studied extensively [16]. Unfortunately, it has been recognized that ideal time-optimal control may be impractical because of the sensitivity to parameter variations, and unmodeled dynamics [17]. To address this issue, a concept of proximate time-optimal (PTO) control has been proposed [18],[21] and successfully applied in, for example, disk-drive head positioning. The main underlying idea considers saturating the control action to facilitate near-time-optimal response to large-signal disturbances and smoothly switching the controller to a standard continuous-time control action in the vicinity of steady state. The power switch, Q1, is an n-channel MOSFET. The diode, CR1, is usually called the catch diode, or freewheeling diode. The inductor, L, and capacitor, C, make up the output filter. The capacitor ESR, RC, (equivalent series resistance) and the inductor DC resistance, RL , are included in the analysis. The resistor, R, represents the load seen by the power stage output.

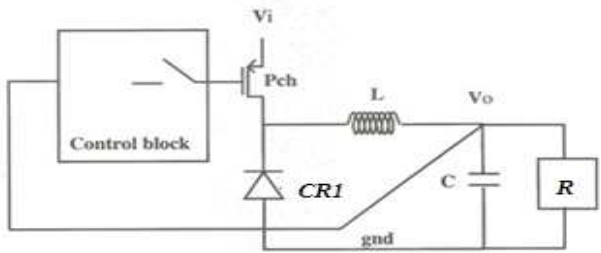

Fig.1.1 Basic Buck Topology

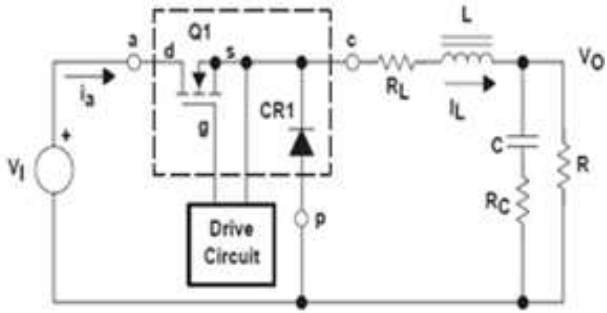

Fig.1.2 Buck power stage Schematic

## 2. Design Consideration

### 2.1 Buck converter steady-state

a) Normal continuous conduction mode (CCM) analysis

The following is a description of steady-state operation in continuous conduction mode [10]. The main result of this section is a derivation of the voltage conversion relationship for the continuous conduction mode buck power stage. This result is important because it shows how the output voltage depends on duty cycle and input voltage or, conversely, how the duty cycle can be calculated based on input voltage and output voltage. Steady-state implies that the input voltage, output voltage, output load current, and duty-cycle are fixed and not varying. Capital letters are generally given to variable names to indicate a steady-state quantity.

In continuous conduction mode, the Buck power stage assumes two states per switching cycle [2][3]. The ON state is when Q1 is ON and CR1 is OFF. The OFF state is when Q1 is OFF and CR1 is ON. A simple linear circuit can represent each of the two states where the switches in the circuit are replaced by their equivalent circuits during each state. The circuit diagram for each of the two states is shown in Fig. 2.1.

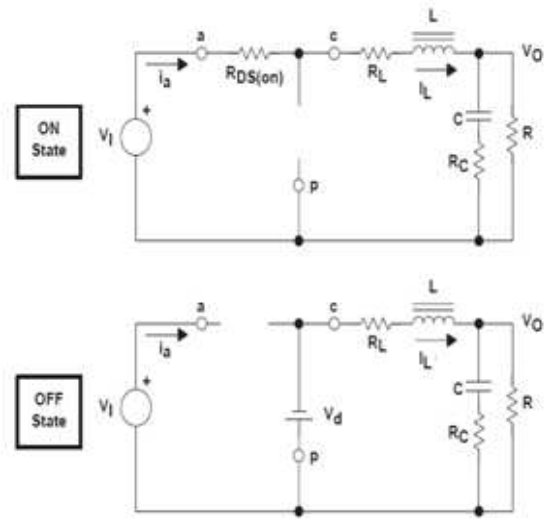

Fig.2.1 Buck Power state states

The duration of the ON state is $D \times T_S = T_{ON}$ where D is the duty cycle, set by the control circuit, expressed as a ratio of the switch ON time to the time of one complete switching cycle, $T_S$. The duration of the OFF state is called $T_{OFF}$. Since there are only two states per switching cycle for continuous mode, $T_{OFF}$ is equal to $(1 - D) \times T_S$. The quantity (1–D) is sometimes called D'. These times are shown along with the waveforms in Fig 2.2.

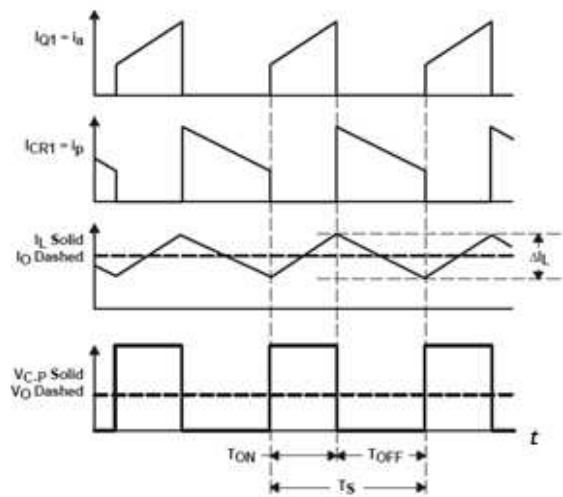

Fig.2.2 Continuous-Mode Buck Power Stage Waveforms

Referring to Fig.2.1, during the ON state, Q1 presents a low resistance, $R_{DS}(ON)$, from its drain to source and has a small voltage drop of $V_{DS} = I_L \times R_{DS}(ON)$. There is also a small voltage drop across the dc resistance of the inductor equal to $I_L \times R_L$. Thus, the input voltage, $V_I$, minus losses,

$(V_{DS} + I_L \times R_L)$, is applied to the left-hand side of inductor, L. CR1 is OFF during this time because it is reverse biased. The voltage applied to the right hand side of L is simply the output voltage, $V_O$. The inductor current, $I_L$, flows from the input source, $V_I$, through Q1 and to the output capacitor and load resistor combination.

During the ON state, the voltage applied across the inductor is constant and equal to $V_L = V_I - V_{DS} - I_L \times R_L - V_O$.

Adopting the polarity convention for the current $I_L$ shown in Fig.2.1, the inductor current increases as a result of the applied voltage. Also, since the applied voltage is essentially constant, the inductor current increases linearly. This increase in inductor current during $T_{ON}$ is illustrated in Fig.2.2. The amount that the inductor current increases can be calculated by using a version of the familiar relationship:

$$V_L = L \times \frac{dI_L}{dt} \implies \Delta I_L = \frac{V_L}{L} \times \Delta_T$$

The inductor current increase during the ON state is given by:

$$\Delta I_L(+) = \frac{(V_I - V_{DS} - I_L \times R_L) - V_O}{L} \times T_{ON}$$

(1)

This quantity, $\Delta I_L(+)$, is referred to as the inductor ripple current.

Referring to Fig.2.1, when Q1 is OFF, it presents a high impedance from its drain to source. Therefore, since the current flowing in the inductor L cannot change instantaneously, the current shifts from Q1 to CR1. Due to the decreasing inductor current, the voltage across the inductor reverses polarity until rectifier CR1 becomes forward biased and turns ON. The voltage on the left-hand side of L becomes $-(V_d + I_L \times R_L)$ where the quantity, $V_d$, is the forward voltage drop of CR1. The voltage applied to the right hand side of L is still the output voltage, $V_O$. The inductor current, $I_L$, now flows from ground through CR1 and to the output capacitor and load resistor combination. During the OFF state, the magnitude of the voltage applied across the inductor is constant and equal to $(V_O + V_d + I_L \times R_L)$. Maintaining our same polarity convention, this applied voltage is negative (or opposite in polarity from the applied voltage during the ON time). Hence, the inductor current decreases during the OFF time. Also, since the applied voltage is essentially constant, the inductor current decreases linearly. This decrease in inductor current during TOFF is illustrated in Fig.2.2. The inductor current decrease during the OFF state is given by:

$$\Delta I_L(-) = \frac{V_O + (V_d + I_L \times R_L)}{L} \times T_{OFF}$$

(2)

This quantity $\Delta I_L(-)$ is also referred to as the inductor ripple current. In steady state conditions, the current increase, $\Delta I_L(+)$, during the ON time and the current decrease during the OFF time, $\Delta I_L(-)$, must be equal. Otherwise, the inductor current would have a net increase or decrease from cycle to cycle which would not be a steady state condition. Therefore, these two equations (1) and (2) can be equated and solved for $V_O$ to obtain the continuous conduction mode buck voltage conversion relationship.

Solving for $V_O$:

$$V_O = (V_I - V_{DS}) \times \frac{T_{ON}}{T_{ON} + T_{OFF}} - V_d \times \frac{T_{OFF}}{T_{ON} + T_{OFF}} - I_L \times R_L$$

(3)

And, substituting $T_S$ for $T_{ON} + T_{OFF}$, and using $D = \frac{T_{ON}}{T_S}$ and $(1 - D) = \frac{T_{OFF}}{T_S}$, the steady-state equation for $V_O$ is:

$$V_O = (V_I - V_{DS}) \times D - V_d \times (1 - D) - I_L \times R_L$$

(4)

A common simplification is to assume $V_{DS}$, $V_d$ and $R_L$ are small enough to ignore. Setting $V_{DS}$, $V_d$ and $R_L$ to zero, the above equation (4) simplifies considerably to:

$$V_O = V_I \times D$$

(5)

b) Normal Discontinuons Conduction Mode (DCM) Analysis

To begin the derivation of the discontinuous conduction mode buck power stage voltage conversion ratio, observe that there are three unique states that the power stage assumes during discontinuous current mode operation [2]. The ON state is when Q1 is ON and CR1 is OFF. The OFF state is when Q1 is OFF and CR1 is ON. The IDLE state is when both Q1 and CR1 are OFF. The first two states are identical to those of the continuous mode case and the circuits of Fig.2.1 are applicable except that $T_{OFF} \neq (1 - D) \times T_S$. The remainder of the switching cycle is the IDLE state. In addition, the dc resistance of the output inductor, the output diode forward voltage drop, and the power MOSFET ON-state voltage drop are all assumed to be small enough to omit. The duration of the ON state is $T_{ON} = D \times T_S$ where D is the duty cycle, set by the control circuit, expressed as a ratio of the switch ON time to the time of one complete switching cycle, Ts. The duration of the OFF state is $T_{OFF} = D_2 \times T_S$. The IDLE time is the remainder of the switching cycle and is given as $T_S - T_{ON} - T_{OFF} = D_3 \times T_S$. These times are shown with the waveforms in Fig.2.3.

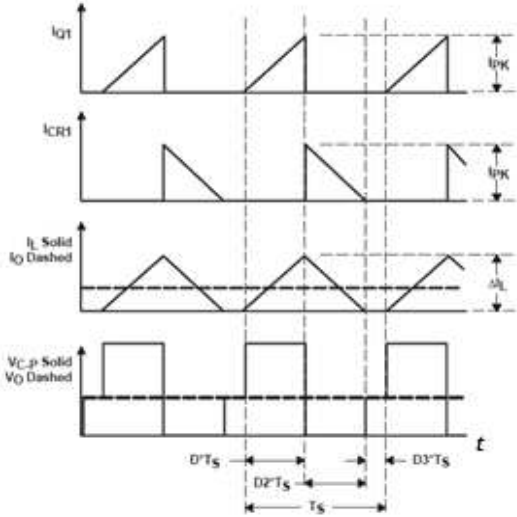

Fig.2.3 Discontinuous-Mode Buck Power Stage Waveforms

Without going through the detailed explanation as before, the equations for the inductor current increase and decrease are given below. A common simplification is to assume $V_{DS}$, $V_d$ and $R_L$ are small enough to ignore. Setting $V_{DS}$, $V_d$ and $R_L$ to zero. The inductor current increase during the ON state is given by:

$$\Delta I_L(+) = \frac{(V_I - V_O)}{L} \times T_{ON} = \frac{(V_I - V_O)}{L} \times D \times T_S = I_{PK}$$

(6)

The ripple current magnitude, $\Delta I_L(+)$ is also the peak inductor current, $I_{PK}$, because in discontinuous mode, the current starts at zero each cycle.

The inductor current decrease during the OFF state is given by:

$$\Delta I_L(-) = \frac{V_O}{L} \times T_{OFF}$$

(7)

As in the continuous conduction mode case, the current increase, $\Delta I_L(+)$ during the ON time and the current decrease during the OFF time, $\Delta I_L(-)$ are equal.

Therefore, these two equations (6) and (7) can be equated and solved for $V_O$ to obtain the first of two equations (6) and (7) to be used to solve for the voltage conversion ratio:

$$V_O = V_I \times \frac{T_{ON}}{T_{ON} + T_{OFF}} = V_I \times \frac{D}{D + D_2}$$

(8)

Now we calculate the output current (the output voltage $V_O$ divided by the output load R). It is the average of the inductor current.

$$I_O = I_{L(avg)} = \frac{V_O}{R} = \frac{I_{PK}}{2} \times \frac{D \times T_S + D_2 \times T_S}{T_S}$$

(9)

Now, substitute the relationship for $I_{PK}$ into the above equation (6) and (9) to obtain:

$$I_O = \frac{V_O}{R} = (V_I - V_O) \times \frac{D \times T_S}{2 \times L} \times (D + D_2)$$

(10)

We now have two equations, the one for the output current just derived and the one for the output voltage (above), both in terms of $V_I$, D, and $D_2$. We now solve each equation (8) and (10) for $D_2$ and set the two equations equal to each other. Using the resulting equation, an expression for the output voltage $V_O$, can be derived. The discontinuous conduction mode buck voltage conversion relationship is given by:

$$V_O = V_I \times \frac{2}{1 + \sqrt{1 + \frac{4 \times K}{D^2}}}$$

(11)

Where K is defined as:

$$K = \frac{2 \times L}{R \times T_S}$$

The above relationship shows one of the major differences between the two conduction modes. For discontinuous conduction mode, the voltage conversion relationship is a function of the input voltage, duty cycle, power stage inductance, the switching frequency and the output load resistance while for continuous conduction mode, the voltage conversion relationship is only dependent on the input voltage and duty cycle.

It should be noted that the buck power stage is rarely operated in discontinuous conduction mode in normal situations, but discontinuous conduction mode will occur anytime the load current is below the critical level.

## 2.2 Burst mode operation

The buck converter is with a burst–mode control method [22][23]. It means the MOSFET can be completely off for one or more switching cycles. The output voltage is regulated by the overall duration of dead time or non–dead time over a number of switching cycles. This feature offers advantages on saving energy in standby condition since it can reduce the effective duty cycle dramatically. In flyback topology, the circuit is mainly designed for discontinuous conduction mode (DCM) in which the inductor current reaches zero in every switching cycle. The DCM burst–mode waveform can be represented in Fig.2.4. It is similar to the pulse–width modulation (PWM) one. In non–isolated topologies such as buck, the circuits are mainly designed for CCM. The CCM burst–mode waveform is different to the PWM waveform in Fig.2.5. Because of this characteristic, burst mode requires a higher peak value of the inductor current in order to have the same level of averaged inductor current (or output current). By using burst mode, the chip will only switch the power devices when the output voltage needs it to do so. During normal DCM mode, the switches will be turned on and off at every cycle. In burst mode, the switches will only be turned on or off when needed. The efficiency can be Improved a lot especially at very light load. As shown in Fig.2.4 and Fig.2.5 burst–mode control produces low–frequency waveform comparing to the switching frequency. Part of the power loss in this low frequency becomes audible noise. Therefore, burst–mode control is not suitable for high power applications such as more than 20 W.

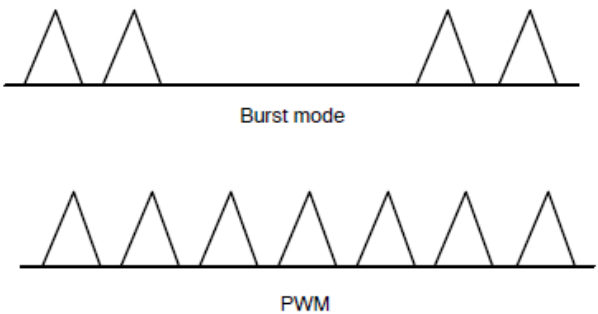

Fig.2.4 DCM Inductor Currents in Burst Mode and PWM Control

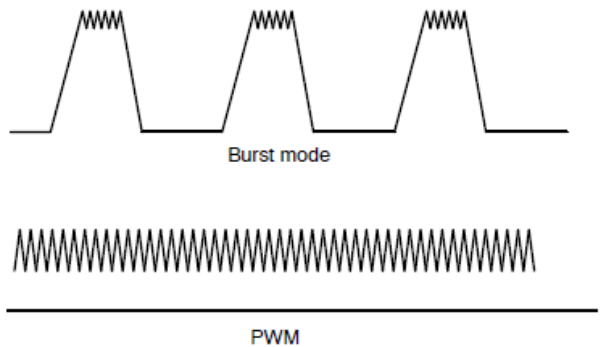

Fig.2.5 CCM Inductor Currents in Burst Mode and traditional PWM Control

Figure 2.6 shows the different inductor voltage and inductor current of normal CCM, DCM and burst conduction mode.

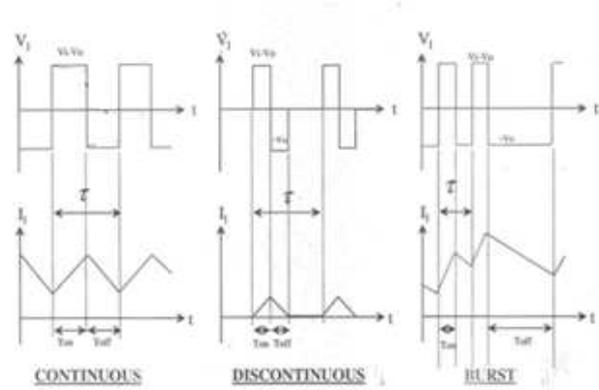

Fig.2.6 continuous, discontinuous and burst conduction mode

## 3. Synchronous and Asynchronous buck dc-dc converter

In this module, we will learn the key differences between synchronous and asynchronous buck topologies, their advantages, disadvantages and their application considerations.

### 3.1 Asynchronous Buck Topology

A typical asynchronous buck DC-DC converter circuit is as shown in the Fig.1.2 [2],[4]. Q1 denotes a MOSFET being used in the top side with a diode CR1 in the bottom side. These are the two main switches that control power to the load. When the MOSFET is turned ON, $V_i$ charges the inductor 'L', capacitor 'C' and supplies the load current. Upon reaching its set output voltage the control circuitry turns OFF the MOSFET (hence called a switching MOSFET). Switching OFF the MOSFET disrupts the current flowing through the inductor. With no path for the current, the inductor will resist this change in the form of a catastrophic voltage spike. To avoid this spike when the MOSFET is turned OFF, a path is provided for the inductor current to continue flowing in the same direction as it did before. This is created by the diode CR1. When the MOSFET turns OFF, the inductor voltage reverses its polarity forward biasing the diode CR1 ON, allowing the current to continue flowing through it in the same direction. When current flows in the diode, it is also known as being in freewheel mode. When the output voltage drops below the set point, the control will turn ON the MOSFET and this cycle repeats to regulate the output voltage to its set value.

### 3.2 Synchronous Buck Topology

The synchronous topology is depicted in the Fig.1.2 and The diode 'CR1' has been replaced with another

MOSFET Q2 [5],[6],[7]. The Q2 is referred to as the synchronous MOSFET while the Q1 is called the switching/control MOSFET. In steady state, the Q2 is driven such that it is complimentary with respect to the Q1. This means whenever one of these switches is ON, the other is OFF. In steady state conditions, this cycle of turning the Q1 and Q2 MOSFETs ON and OFF complimentary to each other regulates $V_o$ to its set value.observe that the Q2 MOSFET will not turn ON automatically. This action needs additional MOSFET drive circuitry within the control IC to turn ON and OFF as needed. Compare this to asynchronous topology where the polarity reversal across the inductor automatically forward biases the diode, completing the circuit.

A. Advantages and disadvantages of these two topologies

The asynchronous topology uses just one MOSFET for the top side as the control switch. There is no so-called shoot-through issue. The IC tends to be smaller and relatively inexpensive. The use of an external diode in most cases eliminates the need for an expensive thermally enhanced IC package to dissipate the heat arising during the freewheel mode. However, the package still needs to take care of heat dissipation arising from switching losses associated with the Q1 MOSFET.

In the synchronous topology the Q2 MOSFET's lower resistance from drain to source $R_{DS(ON)}$ helps reduce losses significantly and therefore optimizes the overall conversion efficiency. However, all of this demands a more complicated MOSFET drive circuitry to control both the switches. Care has to be taken to ensure both MOSFETs are not turned on at the same time. If both MOSFETs are turned on at the same time a direct short from $V_I$ to ground is created and causes a catastrophic failure. Ensuring this direct short, which is also called cross-conduction or shoot-through, does not occur requires more complexity and cost within the IC.

## 4. EXPERIMENTAL RESULT

The circuit has been designed using a standard 0.18-um CMOS technology with dual poly and 5 metals. The supply voltage ranges from 2.6 to 4.2V. The nominal switching frequency is 50 MHz. Experimental results show that, at 2.6-V minimum supply, the output regulation range is 0.5 V – 2.45 V with output current up to 300 mA. The power efficiencies versus current load are shown in Fig.3.1 and Fig.3.2. Higher supply causes larger dynamic dissipation and this worsens the efficiency at low currents. However, the low power of the control (only 35 uA), sustains the overall efficiency at almost one decade below the peak. Fig. 3.4 shows the different output and input of the proposed buck converter, load regulation measurement with output current, ILoad output of the driver, switched from 0 A to 300 mA by on-off current control on PCB. The load switching speed is 2 us.

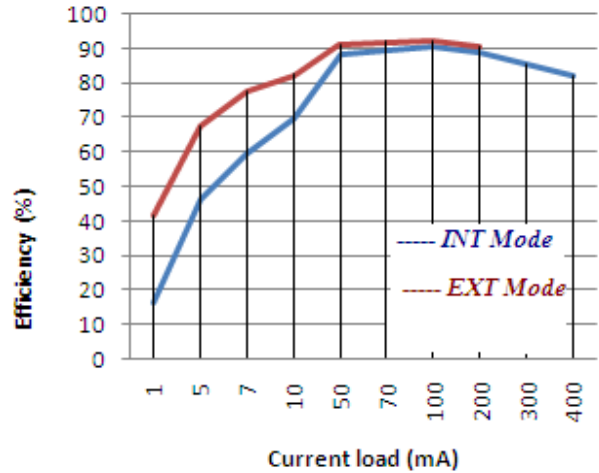

Fig.3.1 efficiency in mode on versus current load

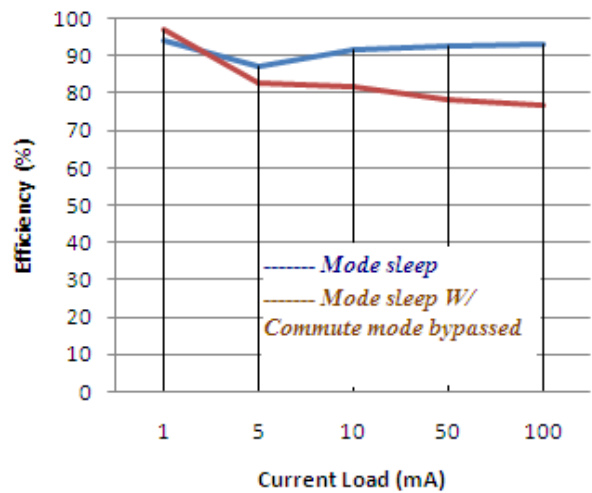

Fig.3.2 efficiency in mode sleep versus current load

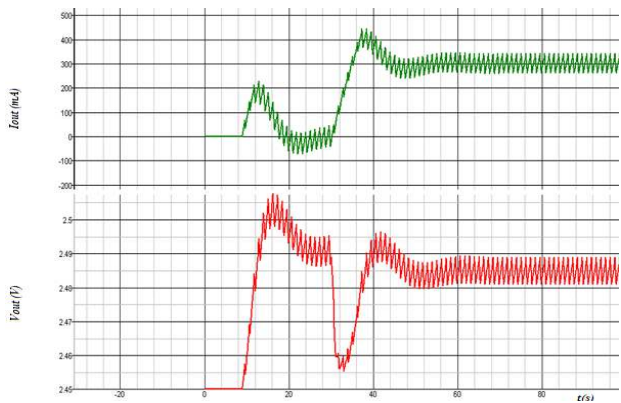

Fig.3.3 output of the proposed buck converter

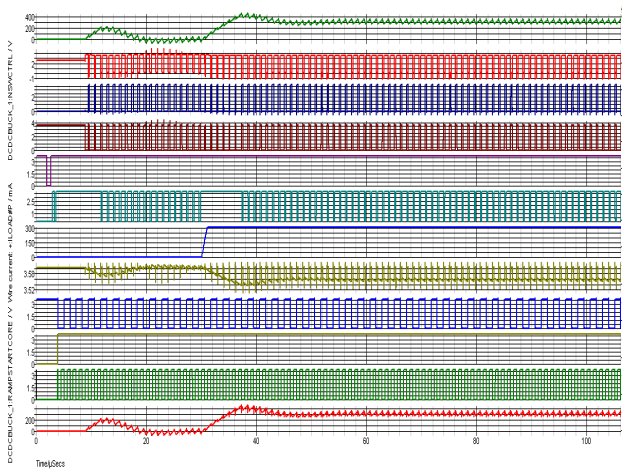

Fig.3.4 simulation results of the proposed buck converter

## 5. CONCLUSION

In this paper, a switched simulation model of the buck converter is presented. We simulate a buck converter by using the software SIMetrix including Power System Blocs and We have realized a novel PWM circuit that can be used to provide a range of Vdd levels for a variety of loads to optimize the efficiency of the converter in mode on and mode sleep. With the use of a deferent value of current load and inner feedback loop between the output and input of the power transistors, we are able to ensure real-time zero voltage switching. This enables the reduction of power consumed by these transistors and achieves power efficiencies over 90% for a variety of loads. Also, an outer feedback loop is employed in the PWM circuit to track the reference voltage level.

**REFERENCES**

[1] Simpson. (2003) Characteristics of Rechargeable Batteries. National.com.[Online].Available: http://www.national.com/appinfo/ power/files/fv.pdf.
[2] Application Report Understanding buck power stages in switch mode power supplies TI literature number slva057
[3] R. W. Erickson, Fundamentals of Power Electronics, New York: Chapman and Hall, 1997.
[4] 380kHz,2A Asynchronous DC-DC Buck Converter ,BCD Semiconductor Manufacturing Limited, Feb. 2011 http://www.bcdsemi.com.
[5] Application Report Designing Fast Response Synchronous Buck Regulators Using theTPS5210, TI Literature Number SLVA044.
[6] Vahid Yousefzadeh, Amir Babazadeh, , Bhaskar Ramachandran, Eduard Alarcón, Pao, , and Dragan Maksimovic´,Proximate "Time-Optimal Digital Control for Synchronous Buck DC–DC Converters", IEEE TRANSACTIONS ON POWER ELECTRONICS, VOL. 23, NO. 4, JULY 2008

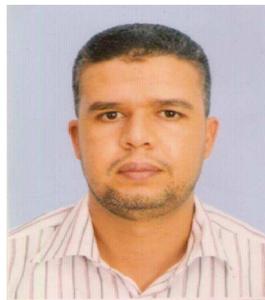

[7] Vahid Yousefzadeh, Dragan Maksimovic´"Sensorless Optimization of Dead Times in DC–DC Converters With Synchronous Rectifiers", IEEE TRANSACTIONS ON POWER ELECTRONICS, VOL. 21, NO. 4, JULY 2006.
[8] P. Gupta and A. Patra, "Super-stable energy based switching control scheme for dc–dc buck converter circuits," in Proc. IEEE ISCAS, 2005, vol. 4, pp. 3063–3066.
[9] K. K. S. Leung and H. S. H. Chung, "Derivation of a second-order switching surface in the boundary control of buck converters," IEEE Power Electron. Lett., vol. 20, no. 2, pp. 63–67, Jun. 2004.
[10] K. S. Leung and H. S. H. Chung, "A comparative study of the boundary control of buck converters using first- and second-order switching surfaces -Part I: Continuous conduction mode," in Proc. IEEE PESC, 2005, pp. 2133–2139.
[11] K. S. Leung and H. S. H. Chung, "A comparative study of the boundary control of buck converters using first- and second-order switching surfaces–Part II: Discontinuous conduction mode," in Proc. IEEE PESC, 2005, pp. 2126–2132.
[12] M. Ordonez, M. T. Iqbal, and J. E. Quaicoe, "Selection of a curved switching surface for buck converters," IEEE Trans. Power Electron., vol. 21, no. 4, pp. 1148–1153, Jul. 2006.
[13] J. T. Mossoba and P. T. Krein, "Exploration of deadbeat control for dc–dc converters as hybrid systems," in Proc. IEEE PESC, Jun. 2005, pp. 1004–1010.

[14] G. E. Pitel and P. T. Krein, "Trajectory paths for dc–dc converters and limits to performance," in Proc. IEEE COMPEL, 2006, pp. 40–47.
[15] P. T. Krein, "Feasibility of geometric digital controls and augmentation for ultrafast dc–dc converter response," in Proc. IEEE COMPEL, 2006, pp. 48–56.
[16] M. Athans and P. Falb, Optimal Control: An Introduction to The Theory and Its Applications. New York: McGraw-Hill, 1966.
[17] S. I. Zinober and A. T. Fuller, "The sensitivity of nominally timeoptimal control systems to parameter variation," Int. J. Contr., vol. 17, no. 4, pp. 673–703, 1973.
[18] W. S. Newman, "Robust near time-optimal control," IEEE Trans. Automat. Contr., vol. 35, no. 7, pp. 841–844, Jul. 1990.
[19] L. Y. Pao and G. F. Franklin, "Proximate time-optimal control of thirdorder servomechanisms," IEEE Trans. Automat. Contr., vol. 38, no. 4, pp. 560–580, Apr. 1993.
[20] L. Y. Pao and G. F. Franklin, "The robustness of a proximate timeoptimal controller," IEEE Trans. Automat. Contr., vol. 39, no. 9, pp. 1963–1966, Sep. 1994.
[21] Laorpacharapan and L. Y. Pao, "Shaped time-optimal feedback control for disk-drive systems with back-electromotive force," IEEE Trans. Magn., vol. 40, no. 1, pp. 85–96, Jan. 2004.
[22] Krit Salah-ddine, Jalal Laassiri and El Hajji Said, "Specification and Verification of Uplink Framework for Application of Software Engineering using RM-ODP", IJCSI International Journal of Computer Science Issues, Vol. 8, Issue 5, No 1, September 2011.
[23] Krit Salah-ddine , Zared Kamal,  Qjidaa Hassan and Zouak Mohcine, "A 100 mA Low Voltage Linear Regulators for Systems on Chip Applications Using 0.18 μm CMOS Technology" IJCSI International Journal of Computer Science Issues, Vol. 9, Issue 1, No 3, January 2012.

**Benlafkih Abdessamad** received the B.S. and M.S. degrees, from university of sciences Dhar El Mehraz Fès. In 2001 and 2003, Respectively. During 2011-2013, He stayed in physics Research Laboratory, faculty of Science University Ibn Tofail Kénitra Morocco to study Analogue Converter DC-DC buck and Boost, mobile satellite communication systems, and wireless access network using CMOS Technology.

**Salah-ddine Krit** received the B.S. and Ph.D degrees in Software Engineering from Sidi Mohammed Ben Abdellah university, Fez, Morroco in 2004 and 2009, respectively. During 2002-2008, He worked as an engineer Team leader in audio and power management of Integrated Circuits (ICs) Research, Design, simulation and layout of analog and digital blocks dedicated for mobile phone and satellite communication systems using Cadence, Eldo, Orcad, VHDL-AMS technology. Currently, He is a professor of Informatics with the Polydisciplinary Faculty of Ouarzazate, Ibn Zohr university in Agadir, Morroco. My research interests include: Wireless Sensor Networks (Software and Hardware), Computer Engineering, Wireless Communications, Genetic Algorithms and Gender and ICT.

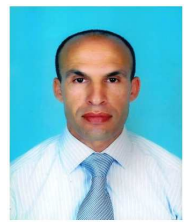